\documentclass[sigconf,authorversion]{acmart}

\AtBeginDocument{%
  \providecommand\BibTeX{{%
    \normalfont B\kern-0.5em{\scshape i\kern-0.25em b}\kern-0.8em\TeX}}}

\setcopyright{rightsretained}
\copyrightyear{2021}
\acmYear{2021}
\acmConference[ARES 2021]{The 16th International Conference on Availability, 
Reliability and Security}{August 17--20, 2021}{Vienna, Austria}
\acmBooktitle{The 16th International Conference on Availability, Reliability and 
Security (ARES 2021), August 17--20, 2021, Vienna, 
Austria}\acmDOI{10.1145/3465481.3465764}
\acmISBN{978-1-4503-9051-4/21/08}

\newcommand{\tabhead}[1]{\textbf{#1}}
 \def\segram{Segram}
\begin{document}

\title{How Private is Android's Private DNS Setting?\\Identifying Apps by Encrypted DNS Traffic}

\author{Michael Mühlhauser}
\email{mm.psi@uni-bamberg.de}
\affiliation{%
  \institution{University of Bamberg}
  \city{Bamberg}
  \country{Germany}
}

\author{Henning Pridöhl}
\email{hp.psi@uni-bamberg.de}
\affiliation{%
  \institution{University of Bamberg}
  \city{Bamberg}
  \country{Germany}
}

\author{Dominik Herrmann}
\email{dh.psi@uni-bamberg.de}
\affiliation{%
 \institution{University of Bamberg}
 \city{Bamberg}
 \country{Germany}}

\begin{abstract}
	DNS over TLS (DoT) and DNS over HTTPS (DoH) promise to improve privacy and security of DNS by encrypting DNS messages, especially when messages are padded to a uniform size.
	Firstly, to demonstrate the limitations of recommended padding approaches, we present \segram{}, a novel app fingerprinting attack that allows adversaries to infer which mobile apps are executed on a device.
	Secondly, we record traffic traces of 118 Android apps using 10 different DoT/DoH resolvers to study the effectiveness of \segram{} under different conditions.
	According to our results, \segram{} identifies apps with accuracies of up to 72\,\% with padding in a controlled closed world setting.
	The effectiveness of \segram{} is comparable with state-of-the-art techniques but \segram{} requires less computational effort.
	We release our datasets and code.
  	Thirdly, we study the prevalence of padding among privacy-focused DoT/DoH resolvers, finding that up to 81\,\% of our sample fail to enable padding.
  	Our results suggest that recommended padding approaches are less effective than expected and that resolver operators are not sufficiently aware about this feature.
\end{abstract}



\maketitle

\section{Introduction}\label{sec:intro}
DNS over HTTPS (DoH)~\cite{hoffman2018dns} and DNS over TLS (DoT)~\cite{hu2016specification} have been proposed to tackle the privacy problems of DNS.
Both protocols provide confidentiality and integrity for DNS by encrypting queries and responses.
Standard organizations have expected traffic analysis attacks on encrypted DNS and thus have taken precautions.\\\\
RFC 8467 outlines recommendations for padding strategies for DNS messages: DNS queries should be padded to multiples of 128 bytes, DNS responses to multiples of 468 bytes~\cite{mayrhofer2018padding}.
Indeed, encryption alone is not enough to prevent traffic analysis attacks: research shows that \emph{websites} can still be identified by encrypted DNS traffic~\cite{houser2019investigation,siby2020encrypted}.
Padding, however, is also not sufficient; it only reduces the accuracy of traffic analysis attacks~\cite{bushart2020padding,houser2019investigation,siby2020encrypted}.
While these studies have shown that websites can be identified by encrypted DNS traffic, previous work neglects \emph{mobile clients}~\cite{bushart2020padding,houser2019investigation,siby2020encrypted}.
In contrast, we focus on the \emph{identification of apps on Android}.
Since Android Pie (2018), Google uses DoT resolvers by default~\cite{kline2018dns}.
That is, Android users are likely representing the largest group, which is using encrypted DNS on a daily basis.

Moreover, previous work has focused on \emph{major} resolvers, e.\,g., Google and Cloudflare~\cite{bushart2020padding,houser2019investigation,siby2020encrypted}.
Some users, however, do not trust large companies; they look for \emph{privacy-oriented} resolvers operated by organizations they trust.
It is an open question whether such resolvers can protect users against traffic analysis attacks to the same extent as large companies.
Consequently, we will not only analyze Google's and Cloudflare's resolvers but also three resolvers operated by non-profit organizations.

The identification of Android apps via traffic analysis infringes users' privacy.
A snapshot of the installed apps can be used to infer user traits such as gender, religion, relationship status, or if the user is a parent~\cite{seneviratne2014predicting,seneviratne2014your}.
These user traits might then be used to create user profiles for targeted advertising~\cite{seneviratne2014your}.
Additionally, the presence of sensitive apps might reveal something about the health status or sexual orientation of users, which might be problematic in some countries.
The main contributions of this paper are:
\begin{enumerate}
    \item We analyze the privacy benefits of encrypted DNS for Android users.
    That is, we propose a novel traffic analysis attack called \emph{\segram{}} and compare its accuracy with several traffic analysis attacks from the website fingerprinting literature.
    Our results for DoH show that \segram{} outperforms related traffic analysis attacks, especially once DNS responses are padded.
    For DoT, \segram{} reaches similar results as state-of-the-art attacks while reducing the runtime for classification substantially.
    \item We collect an extensive dataset for multiple recursive resolvers under several conditions and assess the influence of several factors (e.\,g., padding, app updates, choice of the resolver, caching) on the traffic analysis attacks.
    We are releasing both, our dataset and our implementation of the traffic analysis attacks.\footnote{Code and dataset are available at \url{https://github.com/UBA-PSI/segram}.}
    \item We evaluate the support for DNS padding among an extensive list of recursive resolvers.
    Our results highlight that the majority of recursive resolvers do not enable padding, despite existing guidance on choosing the padding length in RFC 8467.
    Only 36\,\% of DoT, or 19\,\% of DoH resolvers applied padding to DNS responses, although the client requests it.
\end{enumerate}
The rest of the paper is organized as follows:
Section~\ref{sec:bg} reviews DoT and DoH.
In Sect.~\ref{sec:related_work}, we discuss related work, i.\,e., website fingerprinting on DoT and DoH traffic as well as app identification relying on all IP traffic.
Section~\ref{sec:methodology} explains our methodology for app selection, data collection, and the design and evaluation of traffic analysis attacks.
The results are presented in Sect.~\ref{sec:results}.
Additionally, we measure the support for padding among recursive resolvers in Sect.~\ref{sec:measurements}.
Section~\ref{sec:discussion} discusses the results of the traffic analysis attacks and their countermeasures, while we conclude in Sect~\ref{sec:conclusion}.

\section{Background}\label{sec:bg}
The primary purpose of DNS is the translation of domain names into IP addresses.
Typically, there are several parties involved during the name resolution process.
Most operating systems, including Android, implement a local stub resolver, which redirects DNS queries to recursive resolvers and caches received answers.
The client, in our case an Android smartphone, and the recursive resolver exchange DNS packets, typically over UDP, in the clear.
Recursive resolvers might answer DNS queries of the client directly from their cache or by contacting authoritative name servers.

As DNS provides neither confidentiality nor integrity, several privacy and security problems evolve.
Within RFC 7626, DNS is described as possibly one of the weakest links in privacy providing a simple way for surveillance~\cite{bortzmeyer2015dns}.
Besides, research has shown that countries manipulate DNS responses for censorship~\cite{pearce2017global}.
Additionally, users might be tracked based on their DNS traffic~\cite{herrmann2013behaviorbased,kirchler2016tracked,herrmann2016behaviorbased} or they might be redirected to malicious services~\cite{grothoff2018secure}.

DoT and DoH aim to solve some of these problems by encrypting DNS queries and responses, i.\,e., the communication between the client and the recursive resolver.

\emph{DNS over TLS (DoT)} has been standardized in RFC 7858~\cite{hu2016specification}.
It transports DNS messages within TLS connections over port 853.

DNS over HTTPS (DoH) has been standardized by RFC 8484~\cite{hoffman2018dns}.
Unlike DoT, DoH does not transport DNS messages directly within TLS connections but rather embeds DNS messages into HTTPS messages.
DNS messages can either be sent via HTTP GET requests as \emph{base64url}-encoded value for the \textit{dns} parameter or in DNS wire format in the body of HTTP POST requests.
These HTTPS messages are sent over port 443 like normal HTTPS traffic; thus DoH traffic is not easily distinguishable from HTTPS traffic.

\section{Related Work}\label{sec:related_work}
There are numerous studies that try to identify Android apps via traffic analysis~\cite{al-naami2016adaptive,alan2016can,taylor2016appscanner}.
These studies, however, rely on \emph{all traffic} and do not limit themselves to encrypted DNS traffic as we do (see Sect.~\ref{sec:methodology} for the rationale of this restriction).

Al-Naami et al. extract only information from packet headers~\cite{al-naami2016adaptive}.
Among their feature set are packet lengths and bursts, which represent continuous packets in one direction.
The main idea of the authors is to model the dependencies between consecutive bursts.
On a set of 100 financial and social apps, their approach reaches an accuracy of $\approx$ 0.84 with Support Vector Machines.

Alan and Kaur evaluate the applicability of different traffic analysis attacks from website fingerprinting for app identification~\cite{alan2016can}.
The authors consider the 1,595 most popular apps and four different devices.
Apps have been selected from the Play Store on the constraints that they use networking and are compatible with all four devices.
The best approach, which relies on the IP packet size frequency distribution~\cite{herrmann2009website} of the launch time traffic, reaches an accuracy of 0.88.

Taylor et al. implement AppScanner~\cite{taylor2016appscanner}.
Their framework uses, in addition to vectors of raw packet lengths also 54 different statistical measures as features.
The authors show that 110 popular Android apps can be identified with accuracy values of about 0.99.

More relevant for our approach is website fingerprinting based on DoT and DoH~\cite{bushart2020padding,houser2019investigation,siby2020encrypted}.
Houser et al. analyze DoT \cite{houser2019investigation}.
The authors use statistical features from related work \cite{hayes2016kfingerprinting,herrmann2009website,wang2014effective} but also novel features such as the number of DNS messages inside a TLS record.
Additionally, they consider several influencing factors in their evaluation, such as the stability of the results over time and among three recursive resolvers.

Siby et al. use unigrams and bigrams of TLS record sizes and bursts to identify websites by encrypted DNS traffic~\cite{siby2020encrypted}.
Their approach reaches an F1 score of $\approx$~0.90 for unpadded DoH traffic.
If DoH responses are padded, the F1 score decreases to 0.43.
For DoT, the authors evaluate only padded traffic.
In this case, their attack reaches an F1 score of 0.50.
Furthermore, Siby et al. evaluated the robustness over time, across locations, and across infrastructures.

Bushart and Rossow describe a traffic analysis attack based on DNS sequences~\cite{bushart2020padding}.
DNS sequences consist of DNS response sizes in order with the logarithmized time gap in between.
The authors feed these sequences into the k-nearest neighbor classifier, where the Damerau-Levenshtein distance serves as the distance function.
They achieve an accuracy of up to 86\,\% on 9{,}235 websites.

\section{Methodology}\label{sec:methodology}
Given the growing support for encrypted DNS traffic (Android Pie, iOS 14, Firefox 62), we assume that most DNS traffic will be encrypted in the future.
Our goal is, therefore, to identify Android apps using encrypted DNS traffic, i.\,e., assign a \emph{traffic trace} containing DoT or DoH packets to the app that caused it. 

\begin{figure}[tb]
	\centering
	\includegraphics[width=1\linewidth]{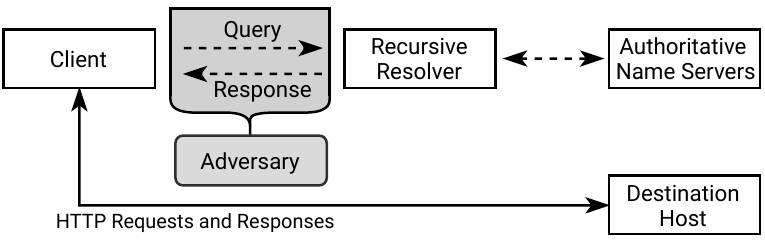}
	\caption{Threat Model for DNS Fingerprinting~\cite{siby2020encrypted}}
	\label{fig:threatmodel}
	\Description{Illustration of our Threat Model: The adversary is located between the user and the recursive resolver. That means, adversaries eavesdrop on DoT/DoH queries and responses only.}
\end{figure}

In our threat model, the adversary is located between the user and the recursive resolver (see Fig.~\ref{fig:threatmodel}) and eavesdrops on the user's network traffic. 

As discussed in Sect.~\ref{sec:related_work}, previous work considered app identification when adversaries can analyze all traffic of a mobile device.
In contrast, we are interested in an adversary with more limited capabilities.
In our scenario, traffic analysis is restricted to DoT/DoH traffic only.
There are various circumstances that are subject to this restriction.
Firstly, our adversary may be located on a segment of the network path between client and recursive resolver that does not have access to the HTTPS traffic (as shown in Fig.~\ref{fig:threatmodel}).
Siby et al. have shown that such adversaries exist~\cite{siby2020encrypted}.
Secondly, adversaries may be interested in performing a cost-efficient analysis, e.\,g., to build advertising profiles for a large number of users.
DoH traces take up 124 times less data than HTTPS traces~\cite{siby2020encrypted}.

\subsection{App and Resolver Selection}
We selected 118 apps (A1–A118) from three different ranking sites: Google Play Store, AppAnnie, and AndroidRank.
For the Google Play store, we have sampled apps randomly from the first twenty entries of the ``top'' and ``top-grossing'' ranking lists of the following categories:
Medical, Dating, Casino, Games, Fitness, News, Lifestyle, and Top~200.
We selected these categories intentionally as the identification of sensitive apps like pregnancy apps or dating apps reveals private information about the users.
For the other two ranking sites, we used similar ranking lists (e.\,g., top usage) inside a category and comparable categories.
After selection and deduplication, apps have been installed from the Google Play Store.

For resolver selection, related work~\cite{bushart2020padding,houser2019investigation,siby2020encrypted} considered mainly Google and Cloudflare since they dominate the market~\cite{lu2019endtoend}.
For the same reason and to compare our results with the related work, we also include them.
As mentioned in Sect.~\ref{sec:intro}, users might prefer resolver operators that claim to be privacy-oriented, e.\,g., non-profit organizations.
We, therefore, also choose three privacy-oriented resolvers operated by non-profit organizations: Quad9, Foundation for Applied Privacy, and Digitale Gesellschaft. All selected resolvers support DoT and DoH.

\subsection{Data Collection}
In our data collection setup, the user's smartphone is connected to a WiFi access point.
The access point captures the network traffic of the smartphone to simulate an adversary.
To set resolvers and start apps, the smartphone is instrumented using Android-Debug-Bridge (ADB) commands.
In our experiments, we use a Google Pixel 3a XL running Android 10.

Our main dataset (D1) consists of 51 traffic traces per app and resolver recorded between April 23, 2020 and May 19, 2020, resulting in 60{,}180 traces in total.
One trace represents the launch of an app for one recursive resolver, e.\,g., the trace with trace\_id 937 represents the start of the Twitter app for Google's DoT resolver on the 28th of April 2020 at 06:23:29.
To record a trace, we proceeded as follows:

\begin{enumerate}
	\item We set the recursive resolver on the smartphone.
	For DoT, we make use of Android's build-in DoT support.
	For DoH, we use Google's Intra app to query DoH resolvers.
	DoT and DoH queries are always padded to 128 bytes, i.\,e., resolvers should pad responses as recommended by RFC 8932~\cite{dickinson2020recommendations}.

	\item The DNS cache is cleared.

	\item We start capturing the network traffic on the access point.

	\item The app is started and we wait 20 seconds for all the resources to load.

	\item The app is closed and capturing is stopped.

	\item The trace is stored together with the used recursive resolver, app, and label.
	
\end{enumerate}
To record the trace of the next app, we go back to Step 2.
Once we have iterated through all apps for one recursive resolver, we change the resolver within the first step and collect traces for all apps on the next resolver.

\begin{figure}[t]
	\centering
	\includegraphics[width=1\linewidth]{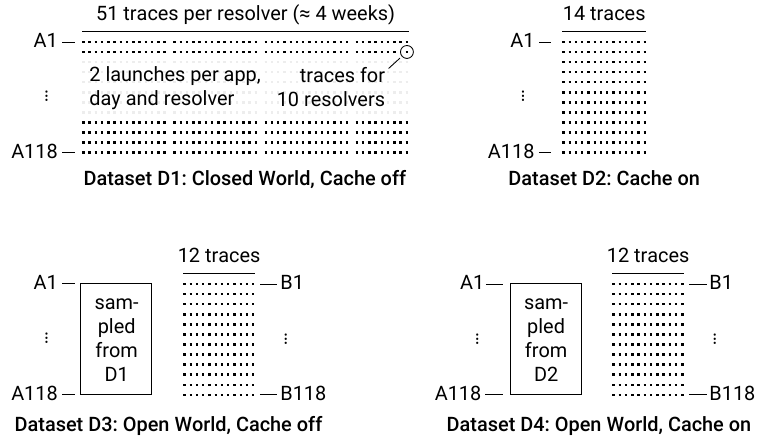}
	\caption{Illustration of our Datasets}
	\label{fig:dataset}
	\Description{Illustration of our Datasets: There are four datasets in total. Two for the closed world scenario, and two for the open world scenario. Within each scenario, one dataset is with caching enabled and one is with caching disabled.}
\end{figure}

Besides the main dataset D1, we have recorded three additional datasets (see Fig.~\ref{fig:dataset}) that will be referenced in upcoming sections.

\subsection{Research Assumptions}
We make several assumptions, partly inspired by similar experiments on website fingerprinting~\cite{herrmann2009website,juarez2014critical,wang2016realistically}:

\begin{itemize}
\item We assume that the adversary has access to the \emph{same device type and operating system} with the same version.
As shown by Alan and Kaur, both properties affect traffic analysis attacks significantly~\cite{alan2016can}.

\item The adversary can use the \emph{same recursive resolver} as the victim, i.\,e., the adversary can reproduce the DNS patterns by querying the same recursive resolvers.
We analyze the effect of this assumption in Sect.~\ref{sec:resolveradapt}.

\item Initially, we operate under the \emph{closed world assumption}, i.\,e., the adversary has a (non-strict) superset of the apps installed on the smartphone that they want to infer and only sees traffic that belongs to an app in this superset.
In practical scenarios, this assumption often does not hold; thus, classifiers often perform worse in practice~\cite{juarez2014critical}.
In Sect.~\ref{sec:ow}, we also analyze an open world scenario, where the adversary has to distinguish whether a traffic trace belongs to an app in their superset or not.

\item We start apps in \emph{isolation}, i.\,e., traffic traces contain only traffic from one app, although background noise may be present.
This assumption is common for traffic analysis research~\cite{juarez2014critical}.
For practical scenarios, there are algorithms to separate overlapping traffic streams~\cite{cui2019revisiting}.

\item We perform \emph{no user interaction} with the app to obtain reproducible results. User interaction may degrade accuracy due to the introduction of unpredictable traffic patterns.

\item DNS caches are cleared before launching an app, i.\,e., we operate in a \emph{cold cache scenario}.
Cold caches allow for better reproducibility and comparison~\cite{herrmann2009website} with related work.
With warm caches, i.\,e., without clearing, we would only observe a (possibly different) subset of DNS queries that an app issues on startup.
We evaluate the effect of warm caches in Sect.~\ref{sec:caching}.

\item We assume the adversary has access to \emph{similar network conditions}, i.\,e., apps operate in a similar network-topological location as the victim.
According to Bushart and Rossow, this assumption is feasible for WiFi networks~\cite{bushart2020padding}.
\end{itemize}

\subsection{DNS Fingerprinting}
We are interested in the effectiveness of app identification using encrypted DNS traffic only, i.\,e., TLS application data of DoT/DoH requests and responses.
To obtain this data, we apply a packet filter, selecting only packets with the expected port numbers, IP addresses of resolvers, and TLS content types.
Thus, the remaining TLS records correspond to DoT/DoH requests/responses.
To distinguish requests and responses, we assign a negative sign to the \emph{requests'} TLS record sizes.
To determine the class of a traffic trace, i.\,e., to identify the app that is represented in that trace, we first transform the traffic trace into a feature vector and subsequently use a Random Forest classifier to obtain its class.
The classifier is trained with labeled traffic traces beforehand.
We define four different types of feature vectors, each corresponding to one traffic analysis attack:

	\paragraph{Frequency Distribution}
	The first feature vector is based on the TLS record size frequency distribution of DoT/DoH packets and is adapted from \cite{alan2016can} and \cite{herrmann2009website}.
	We collect the set of all seen TLS record sizes~\(p_i\) in the training data.
	To obtain a feature vector for a traffic trace, we consider each~\(p_i\) and count the number of TLS records \(\mathrm{\textit{tf}}_{p_i}\) in the trace that match the size~\(p_i\).
	That is, a feature vector is represented by~\(\vec{f} = \langle\mathrm{\textit{tf}}_{p_1}, \mathrm{\textit{tf}}_{p_2}, ..., \mathrm{\textit{tf}}_{p_n}\rangle\)~\cite{herrmann2009website}.

	\paragraph{N-Grams}
	Siby et al. have shown that n-grams are promising for website fingerprinting using encrypted DNS traffic~\cite{siby2020encrypted}.
	We evaluate n-gram features also for app identification.
	Before we describe the feature vector, we define how to construct unigrams and bigrams for TLS record sizes and bursts.
	For a traffic trace, we first obtain an arrival-ordered sequence of all TLS record sizes, i.\,e., the sequence may have duplicates.
	Then, we construct the set of n-grams, e.\,g., for a sequence s \( = [s_1, s_2, s_3, \dots, s_n]\) we obtain the set of unigrams \(\{(s_1), (s_2), (s_3), \dots, (s_n)\}\) and the set of bigrams \(\{(s_1, s_2), (s_2, s_3), \dots, (s_{n-1}, s_n)\}\).
	For bursts, we take the previously mentioned sequence s and add consecutive elements with the same sign, i.\,e., with the same transmission direction, to obtain the burst sequence.
	For example, the sequence [100, 70, 30, -60 -40, 130] is transformed to the burst sequence [200, -100, 130].
	Analogously, we obtain the sets of unigrams and bigrams of the burst sequence.

	The n-gram feature vector is a concatenation of four sub-feature vectors: TLS record-size unigram and bigram, and burst size unigram and bigram.
	Each sub-feature vector contains the frequency of n-gram occurrences, i.\,e., we first determine all available n-grams \(n_i\) of all traffic traces.
	Then, we count the occurrence of each \(n_i\) in an individual traffic trace to obtain its sub-feature vector.
	
	\paragraph{Distances of DNS Sequences}
	Bushart and Rossow used DNS sequences to classify websites using encrypted and padded DNS traffic, obtained from Cloudflare's DoT resolver~\cite{bushart2019padding,bushart2020padding}.
	The authors define DNS sequences as an alternating sequence of DNS response sizes and time gaps between those DNS responses~\cite{bushart2020padding}.
	A time gap \(t\) in ms is encoded with \(\lfloor \log_2{t}\rfloor\) and included in the sequence if $t > 0$.
	We show an exemplary DNS sequence below.
	\begin{displaymath}\label{eq:dnsseq_bushart}
	\mathrm{\textit{DNS}}\_{\mathrm{\textit{Seq}}} = \mathrm{\textit{Msg}}\left(468\right)\; \mathrm{\textit{Gap}}\left(8\right)\;\mathrm{\textit{Msg}}\left(468\right)\; \mathrm{\textit{Gap}}\left(7\right)\;\mathrm{\textit{Msg}}\left(468\right)
	\end{displaymath}
	
	Based on these DNS sequences, Bushart and Rossow classify a traffic trace with a k-nearest neighbor (kNN) classifier.
	The authors use the Damerau-Levenshtein distance between two DNS sequences as custom distance metric, i.\,e., their approach does not use a feature vector but distances between feature vectors~\cite{bushart2020padding}.
	Thus, their approach is computationally expensive.
	The authors specify different costs for operations of the Damerau-Levenshtein distance: insertion, deletion, substitution, and transposition~\cite{bushart2019padding}.

  	We re-implemented Bushart's and Rossow's attack (B\&R attack) to evaluate its effectiveness for app identification, extending the evaluation to multiple resolvers and to DoH traffic.
	We have not tuned any hyper-parameters for the different costs but relied on the parameter values obtained by the authors.

	\paragraph{Segram: N-Grams of DNS Sequences}
	We take the idea of DNS sequences as described above but follow a different approach by turning the DNS sequence into a feature vector for an efficient classifier.
	While the B\&R attack only uses responses, we also include requests in the DNS sequence~\cite{bushart2020padding}.
	We encode time gaps \(t\) in ms as \(\lfloor \log_2{t}\rfloor\) and denote record sizes in bytes with the sign indicating the transmission direction.

	Below is an exemplary DNS sequence for a traffic trace.
	A DoT or DoH request is sent with 154 bytes, followed by a time gap of 274 ms.
	Afterward, the recursive resolver sends the response consisting of 204 bytes.
	Time gaps are only included in the DNS sequence if \(\lfloor \log_{2}{\left(1 + t\right)}\rfloor \ge 5\).
	That is, we exclude more time gaps as in the B\&R attack, which makes the approach less susceptible to network effects~\cite{bushart2020padding}.
	\begin{displaymath}
		\label{eq:dnsseq}
		\mathrm{\textit{DNS}}\_{\mathrm{\textit{Seq}}} = \mathrm{\textit{Msg}}\left(-154\right)\; \mathrm{\textit{Gap}}\left(8\right)\;\mathrm{\textit{Msg}}\left(204\right)
	\end{displaymath}
	
	Analogously to the n-gram feature vector, we construct the feature vector for \segram{} based on the DNS sequence, i.\,e., we obtain the unigrams, bigrams, and trigrams for the DNS sequence and calculate the corresponding frequencies.
	We experimented with several n-grams and found the above combination of unigrams, bigrams, and trigrams most suitable for encrypted and padded DNS traffic.

\subsection{Runtime Benchmarks}
As previously mentioned, the B\&R attack is computationally expensive due to their use of the Damerau-Levenshtein distance and the kNN classifier~\cite{bushart2020padding}.
To quantify the runtime difference between \segram{} and the B\&R attack, we classify 100 randomly selected traffic traces of distinct apps from dataset D1 and train on the remaining ones for each DoT/DoH resolver.
We run each benchmark ten times and report the average runtime and standard deviation.

\segram{} is implemented using Python and scikit-learn and runs on one CPU core.
As our Python implementation of the B\&R attack was too slow to perform the benchmarks within reasonable time, we reimplemented the B\&R attack in Go, with and without parallelization.
For parallelization, the distance calculations to all DNS sequences during the kNN classification run in parallel on all available cores.
The custom Damerau-Levenshtein distance function was implemented using memoization.
We note that kNN classification cannot be optimized using metric trees (such as Ball or KD trees), since DNS sequences are not feature vectors but have variable length.
The benchmarks run on an AMD EPYC 7451 24-core processor with ECC RAM clocked at 2666 MHz.

\section{Results}\label{sec:results}
To measure the effectiveness of the traffic analysis attacks in different settings, we perform stratified 5-fold cross-validation to report accuracy values, i.\,e., the fraction of correctly identified apps.
Unless stated otherwise, we use a sample size of 51 traffic traces per app for cross-validation (see Dataset D1 in Fig.~\ref{fig:dataset}).

\subsection{Overview}
Table 1 shows accuracy values, i.\,e., the fraction of correctly identified apps, for different recursive resolvers and traffic analysis attacks.
We omitted standard deviations for clarity of presentation as values are below 0.012 for \segram{} and below 0.013 for B\&R.

For DoT, resolvers that do not apply padding (AP, Q9, DG) allow for high accuracy values above 0.95 for all performed traffic analysis attacks.
Once padding is applied, e.\,g., for Google (GO) and Cloudflare (CF), the accuracy drops below 0.30 (0.20--0.27), except for the B\&R attack and \segram{}.
For Cloudflare, these two attacks reach an accuracy of 0.72, outperforming the other attacks by 45 percentage points (0.27 for n-grams vs. 0.72 for \segram{}).
For Google, the B\&R attack reaches an accuracy of 0.78, while the accuracy for \segram{} is only 0.67.
\segram{}'s classification runtime, however, is lower (see Sect.~\ref{sec:performance}).
Although the accuracy decreases with padding, \segram{}'s accuracy is still substantially higher than a random guess, which would classify an app correctly with an accuracy of $1/118=0.008$.

\begin{figure}[tb]
	\centering
	\includegraphics[width=.95\linewidth]{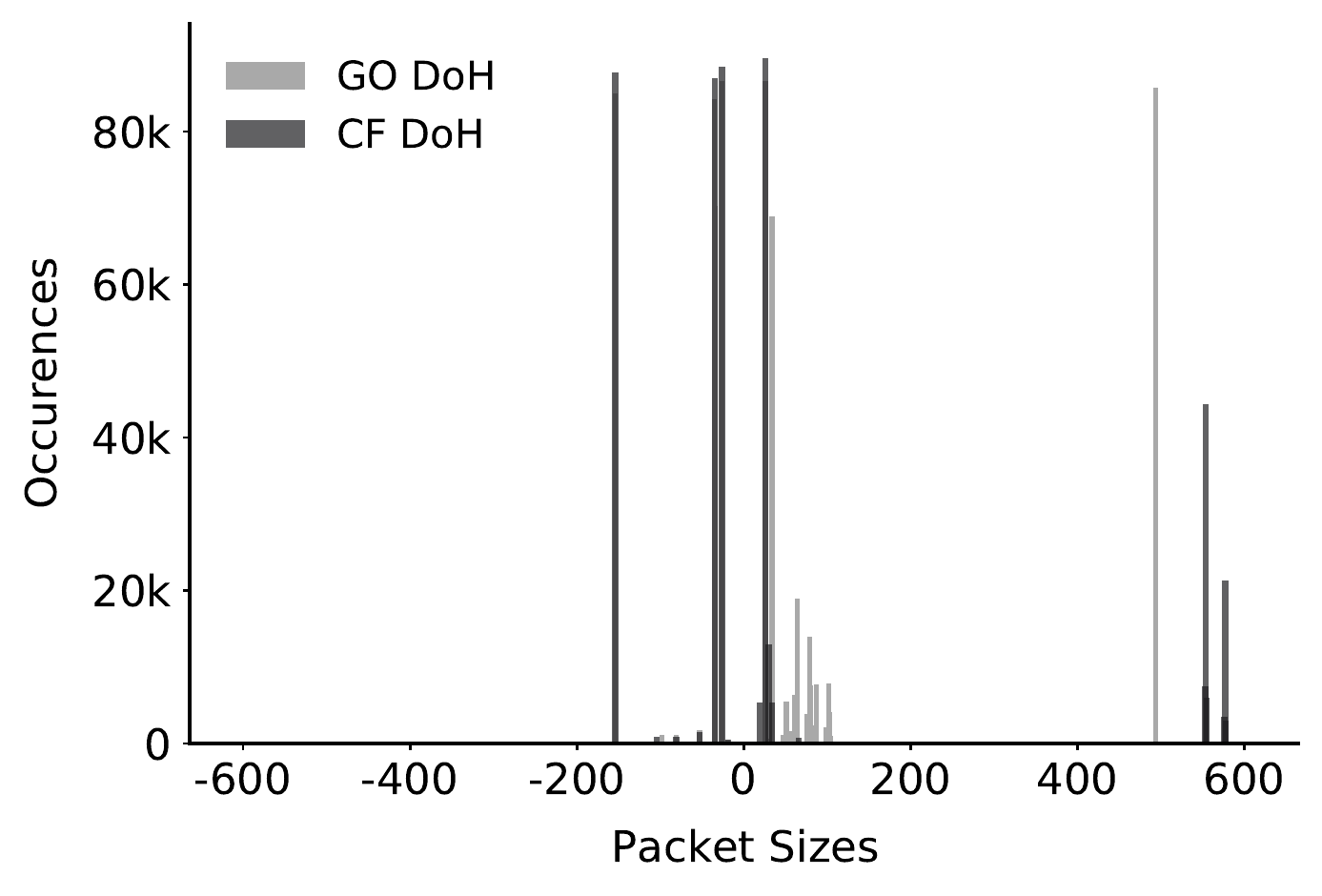}
	\caption{Distribution of packet sizes for Google and Cloudflare; negative packet sizes denote packets sent from the client to the recursive resolver}
	\label{fig:diffgocf}
	\Description{Distribution of packet sizes for Google and Cloudflare.}
\end{figure}

The accuracy values for DoH are above 0.95 when padding is not applied except for the B\&R attack (0.87).
For padding, we see differences between Google and Cloudflare, with Cloudflare having lower accuracy values.
Both resolvers pad responses to 468 bytes, but they differ in the number of packets they transmit and the packet size.
On average, Google exchanges 111 DoH packets, while Cloudflare transmits only 78.
Figure~\ref{fig:diffgocf} shows the distribution of packet sizes for both resolvers.
Observe that only Google sends many packets between 0 and 200 bytes, which might lead to increased accuracy.
For DoH, \segram{} outperforms all the other attacks, including the B\&R attack, albeit only by 8 percentage points for Cloudflare (0.56 vs. 0.64) and 5 percentage points for Google (0.67 vs. 0.72).

\begin{table*}
	\centering
	\caption{Accuracy Values for the Traffic Analysis Attacks\label{tab:results} on Different Resolvers}
	\begin{tabular}{lccccc}
		&&&\multicolumn{1}{c}{\textbf{DoT}}\\
		\cmidrule{2-6}
		& \tabhead{AP} & \tabhead{Q9} & \tabhead{DG} & \tabhead{GO} & \tabhead{CF} \\
		\tabhead{Freq. Distr.}       & 0.99 		& 0.99		& 0.98      & 0.20		& 0.20	\\
		\textbf{N-Grams}            & 0.99      & 0.99      & 0.98      & 0.25      & 0.27	\\
		\textbf{B\&R Attack}			& 0.95		& 0.93		& 0.93		& 0.78		& 0.72	\\
		\textbf{\segram{}}      		& 0.99      & 0.98      & 0.99		& 0.67      & 0.72	\\
	\end{tabular}
	\hspace*{2mm}
	\begin{tabular}{ccccc}
		&&\multicolumn{1}{c}{\textbf{DoH}}\\
		\cmidrule{1-5}
		\tabhead{AP} & \tabhead{Q9} & \tabhead{DG} & \tabhead{GO} & \tabhead{CF} \\
		0.96			& 0.95      & 0.95			& 0.49			& 0.23   \\
		0.97        	& 0.95      & 0.95      	& 0.61          & 0.36   \\
		0.87			& 0.79		& 0.78			& 0.67			& 0.56	 \\
		0.97        	& 0.94		& 0.95      	& 0.72          & 0.64   \\
	\end{tabular}\\[2mm]
	\scriptsize{\textbf{AP} Applied Privacy\quad \textbf{Q9} Quad9\quad \textbf{DG} Digitale Gesellschaft\quad \textbf{GO} Google\quad \textbf{CF} Cloudflare}
\end{table*}

\subsection{Runtime Comparison}\label{sec:performance}
Table~\ref{tab:res_benchmark} shows the runtime for \segram{}, and our implementation of the B\&R attack~\cite{bushart2020padding}.
We report the average runtime for the classification of 100 traffic traces for each DoT and DoH resolver.
The relative standard deviation for the average runtime in our measurements is between 0.4\,\% and 2.1\,\%.

Given that, we are able to make three observations based on our measurements:
Firstly, for most resolvers, the classification of DoH traces with the B\&R attack takes longer than the classification of DoT traces (e.\,g., 203 vs. 1021 seconds for Digitale Gesellschaft).
There is, however, an exception for Applied Privacy and Quad9.
The average runtime for these two resolvers is on the same level for DoT and DoH traces (e.\,g., 54 vs. 56 seconds for Quad9).

Secondly, \segram{}'s runtime is significantly lower than the B\&R attack.
Although we ran the B\&R attack in parallel, \segram{} still outperforms the B\&R attack.
\segram{} is able to classify the 100 traffic traces for all resolvers under 0.04 seconds.
Contrary, the B\&R attack needs 46 seconds in the best case on DoT, or 48 seconds in the best case on DoH.
Thirdly, we observe that although we use a processor with 24 cores, the average runtime for the parallel B\&R attack does only decrease by a factor of about 4.
Further analysis with \emph{perf} showed that our implementation of the B\&R attack is memory-bound and not compute-bound.

However, we have to consider the training time of \segram{}'s Random Forest classifier as there is no training time for the kNN classifier.
Nevertheless, the training time is negligible for practical scenarios, as the Random Forest classifier needs only between six and twelve seconds for the selected resolvers.

\begin{table}[tb]
	\caption{Average Runtime in Seconds for the Classification of 100 Traffic Traces for Segram and the B\&R attack\label{tab:res_benchmark}}
	\begin{tabular}{lccccccc}
		\toprule
		\tabhead{Attack} & \tabhead{Parallel} & \tabhead{Protocol} & \tabhead{AP} & \tabhead{Q9} & \tabhead{DG} & \tabhead{GO} & \tabhead{CF} \\
		\midrule 
		B\&R 	& no   	& DoT   & 203 & 223 & 203 & 178 & 224 \\
		B\&R 	& no   	& DoH   & 193 & 216 & 1021 & 972 & 516 \\
		B\&R 	& yes  	& DoT   & 51 & 54 & 51 & 46 & 55 \\
		B\&R	& yes 	& DoH   & 48 & 56 & 220 & 212 & 122 \\
		Segram 	& no   	& DoT   & 0.03 & 0.03 & 0.03 & 0.03 & 0.03 \\
		Segram 	& no 	& DoH   & 0.04 & 0.04 & 0.04 & 0.04 & 0.03 \\
		\bottomrule
	\end{tabular}
\end{table}

\subsection{Influence of the Sample Size}
\begin{figure}[tb]
	\centering
	\includegraphics[width=.95\linewidth]{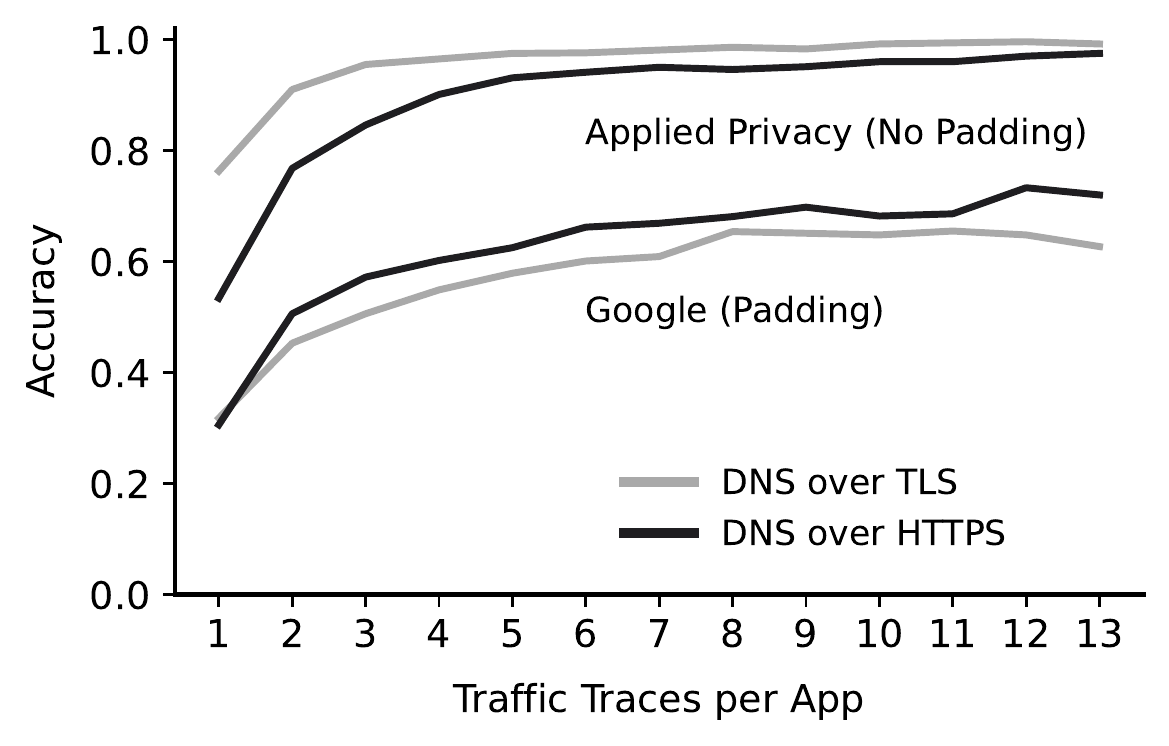}
	\caption{Influence of the Sample Size}
	\label{fig:samplesize}
	\Description{Influence of the Sample Size: Accuracy is increasing with more samples.}
\end{figure}

Collecting one traffic trace for all 118 apps on five resolvers with DoT and DoH takes about 10 hours on our setup.
We, therefore, are interested in how small the sample size for training can be chosen without a significant loss in accuracy.
That is, we vary the number of traces per app from one to thirteen and measure the effect on the classifier's accuracy.
We consider only \segram{} on one recursive resolver without padding (Applied Privacy) and one with padding (Google), see Fig.~\ref{fig:samplesize}.
We chose Google as recursive resolver representing padding since \segram{} performs worst on DoT on that resolver and the difference between \segram{} and the B\&R attack is the smallest on DoH.
Additionally, we consider only traffic traces from one week (14 traces) so that app updates do not influence the results (see Sect.~\ref{sec:updates}).

As shown in Fig.~\ref{fig:samplesize}, the accuracy is increasing steadily for 1 to 5 traces for all resolvers.
With 6 DoT traces per app, we already observe an accuracy of 0.98 without padding.
Once DoT responses are padded, the accuracy decreases to 0.60 with 6 traces and improves to a maximum of 0.66 with 11 traces.

For DoH, accuracy values vary more depending on the number of traces.
Even if responses are not padded, accuracy increases from 0.94 with 6 traces to 0.98 with 13 traces.
With padding, accuracy increases from 0.66 with 6 traces to 0.72 with 13 traces.

\subsection{Influence of Updates}\label{sec:updates}
Updates of apps might lead to changing DNS patterns and thus decreasing accuracy.
To evaluate the need for an up-to-date dataset for an adversary, we installed all available updates at three points in time during data collection.
To be precise, updates have been installed on the 2nd, 9th, and 16th of May 2020.
With the first update, 77 apps were updated, with the second 55, and with the third 38.

We take traces from the first week (25th April -- 2nd May), which serve as a baseline for future updates.
The corresponding accuracy values for the baseline are obtained with cross-validation.
For the classification of traces after the first, second, and third update, we train the classifier with all traces of the baseline.

\begin{figure}[tb]
	\centering
	\includegraphics[width=.95\linewidth]{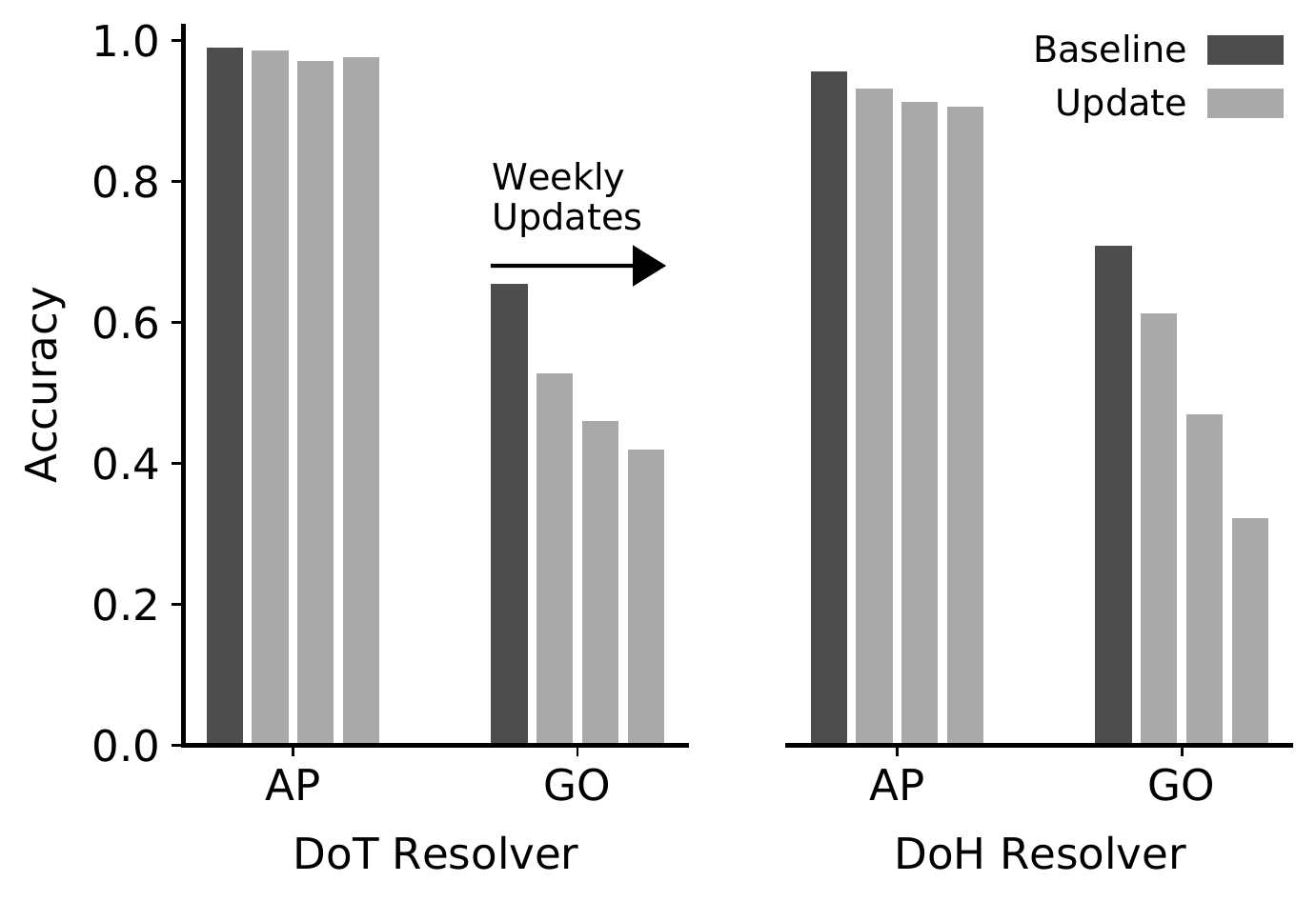}
	\caption{Influence of Updates}
	\label{fig:updates}
	\Description{Influence of Updates: Accuracy decreases substantially after each week for resolvers with padding. Without padding, accuracy is relatively stable.}
\end{figure}

Figure~\ref{fig:updates} shows the results for \segram{} for one recursive resolver without padding (Applied Privacy) and one with padding (Google).
With updates, the accuracy decreases slightly from 99\,\% to 97\,\% for DoT traffic, or from 96\,\% to 91\,\% for DoH traffic.
Once recursive resolvers pad DNS responses, the accuracy drops substantially from 65\,\% to 42\,\% for DoT traffic, or from 71\,\% to 32\,\% for DoH traffic.

\subsection{Predicting on Different Recursive Resolvers}\label{sec:resolveradapt}
Adversaries might want to limit their data collection efforts to only a few resolvers.
We, therefore are interested in whether it is feasible to train on one resolver and classify traffic traces of another resolver.
Again, we only report accuracy values for \segram{}.
For DoT resolvers, the accuracy is generally below 0.2 (mostly even below 0.1), with one exception: If we train on Google and predict on Cloudflare or vice versa, we still reach relatively high accuracies of 0.41 (0.67 or 0.72 when trained on the same resolver).
We remind that both Google and Cloudflare apply padding.

For DoH, we do not observe the same effect on recursive resolvers with padding.
Accuracy is generally below 0.15, with two exceptions.
If we train on Cloudflare, we obtain slightly increased accuracy values for Digitale Gesellschaft (0.30).
For Quad9 and Applied Privacy, we get high accuracy values above 0.93 in both directions.

\subsection{Influence of DNS Caching}\label{sec:caching}
Different apps might resolve the same domains, e.\,g., when loading advertisements.
Thus, some DNS responses might be served from the cache and cannot be observed by an adversary.
Furthermore, users do not open all apps sequentially in the same order as our instrumented smartphone does.

To model the influence of caching, we perform another experiment and open the apps A1–A118 in a different order for every resolver \emph{without clearing caches}. We interrupt data collection briefly after every 16th app to simulate times of inactivity. We choose the interrupt duration randomly from common TTL values observed with the same apps from unencrypted DNS; the maximum TTL was set to 300 seconds, which represents the 75th percentile.

We evaluate caching effects on Dataset D2 (see Fig.~\ref{fig:dataset}).
That is, we collect 14 traces for each app and recursive resolver.
Caching effects are almost negligible for most resolvers, with accuracy values dropping less than 0.05.
For Cloudflare's DoT resolver the accuracy drops from 0.72 to 0.61 (--0.11), though.
This result is in line with studies on website fingerprinting~\cite{bushart2020padding,herrmann2009website}.

\subsection{Open World}\label{sec:ow}
Due to a large number of existing apps, adversaries have to restrict themselves to a limited set of apps they want to identify.
We consider three sets: \emph{monitored apps} that the adversary wants to identify, \emph{unmonitored apps} that are used only for training to represent all other apps, and \emph{unknown apps} that are not monitored and only used for testing.
For instance, if an adversary wants to block 20 apps in her network, these 20 apps would be considered as monitored apps.

In this scenario, the adversary's classifier should output either that an app \emph{is monitored} or not; therefore the adversary selects randomly 100 apps from the Google Play Store and labels them as unmonitored (these 100 apps are disjoint with the monitored apps).
In the real world, the classifier likely sees apps, which are neither part of the monitored apps nor the unmonitored apps, i.\,e., apps which are not part of the 120 selected apps.
To measure how the classifier deals with these previously unseen apps, we consider also a set of unknown apps.

We report results for the binary case, i.\,e., the adversary decides only whether a traffic trace belongs to any monitored app or represents an unknown app~\cite{juarez2014critical,wang2014effective}.
Furthermore, we also evaluate \segram{} in the multi-class case, i.\,e., the adversary decides not only whether an app is monitored but also which app is represented by that specific trace~\cite{panchenko2016website, wang2014effective}.

We select 118 further apps B1–B118 from the same sensitive categories as in the closed world to evaluate \segram{} in the open world scenario (see Dataset D3 in Fig.~\ref{fig:dataset}); thus, we have 236 apps in total.
Then, our training and test dataset is created by randomly selecting apps out of all 236 apps.
Table~\ref{tab:dataow} shows how many apps and traces are selected for the monitored, unmonitored, and unknown set.
The evaluation in the binary case and in the multi-class case is based on the same dataset; only the class labels are different.

\begin{table}[tb]
	\centering
	\caption{Dataset for Open World Evaluation}
	\label{tab:dataow}
	\begin{tabular}{lcccc}
		\toprule
		\tabhead{Type}    & \multicolumn{2}{c}{\tabhead{Training Data}}    &  \multicolumn{2}{c}{\tabhead{Test Data}}\\
		\cmidrule(l){2-3} \cmidrule(l){4-5}
		& \tabhead{Apps} & \tabhead{Traces p. App}    & \tabhead{Apps} & \tabhead{Traces p. App} \\
		\midrule
		Unknown           &        --      & --        &    100 	& 12    \\
		Monitored         &        10      & 30        &    10      & 10    \\
		Unmonit.       &        100     & 3         &    --      & --    \\
		\bottomrule
	\end{tabular}
\end{table}

\begin{figure*}[t]
	\centering
	\includegraphics[width=0.44\linewidth]{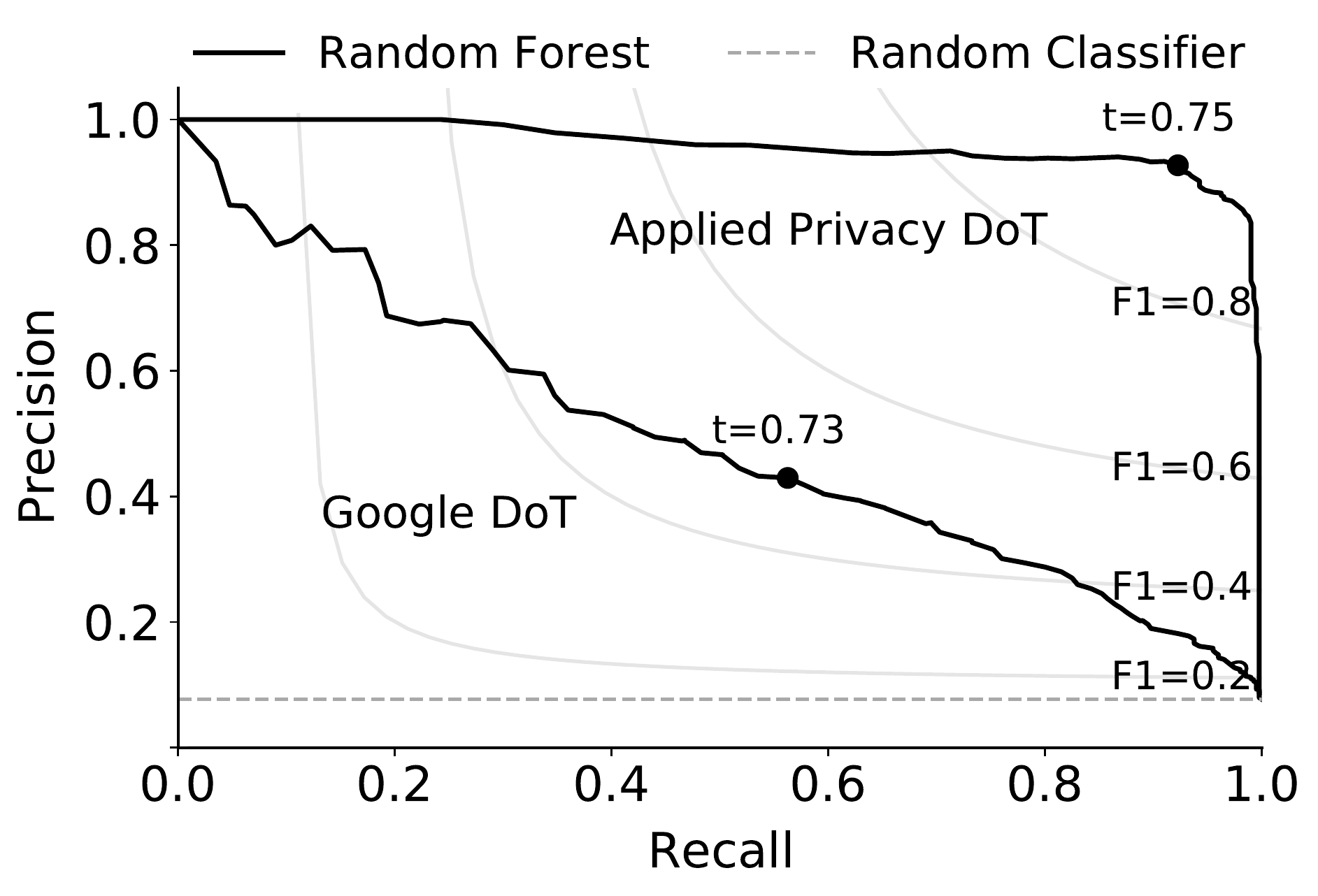}
	\hspace{4mm}
	\includegraphics[width=0.44\linewidth]{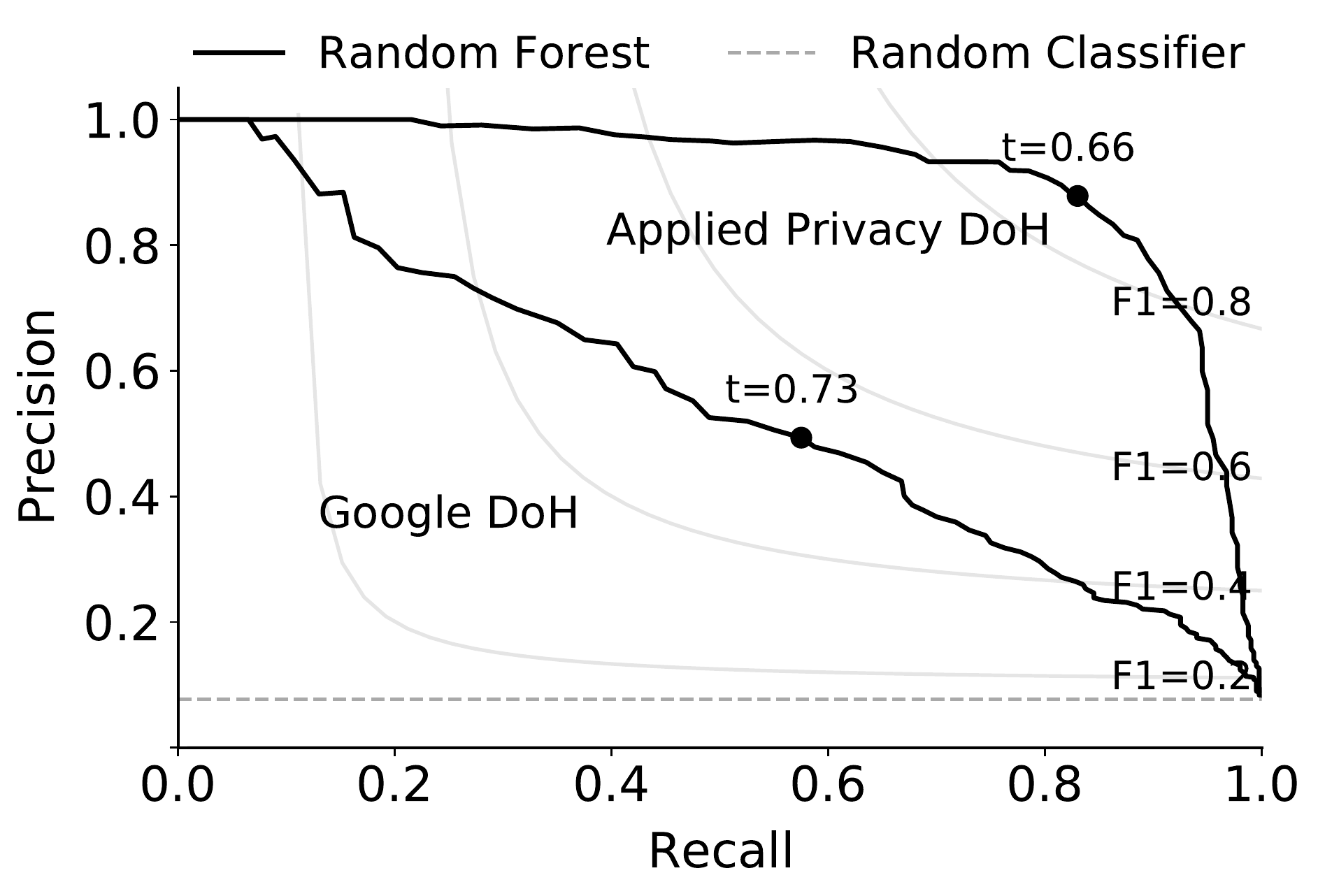}

	\caption{Precision-Recall Curves for \segram{}}
	\label{fig:pr_curves}
	\Description{Precision-Recall Curves for \segram{} for Applied Privacy and Google for DoT and DoH traffic.}
\end{figure*}

\paragraph{Binary Case} 
Similar to the related work~\cite{siby2020encrypted}, we follow the method by Stolerman et al.~\cite{stolerman2014breaking} to evaluate \segram{} in the binary case.
That is, we classify an app as monitored only if the classifier predicts the class as monitored with a probability larger than a certain threshold.
Figure~\ref{fig:pr_curves} illustrates the results for different thresholds; it depicts the average precision-recall curve for four iterations with different monitored apps.

For Applied Privacy’s DoT resolver, we obtain the best F1 score with about 0.92 at threshold \(t = 0.75\).
Results change only slightly on DoH traffic.
The F1 score for Applied Privacy's DoH resolver is about 0.85 at threshold \(t = 0.66\).
With padding, we reach an F1 score of \(\approx 0.49\) for Google's DoT resolver at threshold \(t = 0.73\), or 0.53 at threshold \(t = 0.73\) for Google's DoH resolver.

These F1 scores still pose a threat for users' privacy when comparing the results to a random classifier. 
In fact, a random classifier is represented here by the number of traces of monitored apps divided through the total number of traces.
Figure~\ref{fig:pr_curves} depicts also the performance of the random classifier having an F1 score below 0.20 for all possible thresholds.

\paragraph{Multi-Class Case}
The adversary wants to determine not only whether an app is monitored but also which of the monitored apps is observed in the multi-class case~\cite{panchenko2016website,wang2014effective}.
We also use the dataset from Table~\ref{tab:dataow} to evaluate \segram{}.

The results are shown in Table~\ref{tab:res_ow}.
Note that we report the Macro-F1 score instead of the accuracy as the open world dataset is imbalanced.
While the F1 score for \segram{} is about 0.99 for DoT in the closed world setting, we still reach values between 0.94 and 0.95 in the multi-class case in the open world.
However, if DoT resolvers implement padding, the F1 score drops by 0.10 and 0.14 compared to the closed world.
F1 scores between 0.56 and 0.57 on encrypted and padded DoT traffic are still relatively high in comparison to a random classifier, though.

The results for DoH traffic are affected similarly.
The F1 score decreases by 0.06 for unpadded DoH traffic in comparison to the closed world.
Nevertheless, we still reach F1 scores between 0.87 and 0.92 for unpadded DoH traffic.
If DoH resolvers pad responses, the F1 score decreases by 0.06 in comparison to the closed world (0.54 for Cloudflare vs. 0.66 for Google).

\begin{table}[tb]
	\caption{F1 Scores in the Multi-Class Case\label{tab:res_ow}}
	\begin{tabular}{lcccccc}
		\toprule
		\tabhead{World} & \tabhead{Protocol} & \tabhead{AP} & \tabhead{Q9} & \tabhead{DG} & \tabhead{GO} & \tabhead{CF} \\
		\midrule 
		Open   & DoT   & 0.95 & 0.95 & 0.94 & 0.56 & 0.57 \\
		Closed & DoT   & 0.99 & 0.98 & 0.99 & 0.66 & 0.71 \\
		Open   & DoH   & 0.91 & 0.87 & 0.92 & 0.66 & 0.54 \\
		Closed & DoH   & 0.97 & 0.94 & 0.96 & 0.71 & 0.62 \\
		\bottomrule
	\end{tabular}
\end{table}

\subsection{Open World and Caching}
As shown in the previous section, \segram{} is able to identify apps in the open world setting in the binary and in the multi-class case when caching is disabled.

To evaluate \segram{} in the open world setting \emph{with} caching, we consider Dataset~D4 and sample apps from it (see Fig.~\ref{fig:dataset}).
We still differentiate between monitored, unmonitored, and unknown apps.
Table~\ref{tab:dataowcache} shows how many apps and traces were selected for the respective sets.

\begin{table}[tb]
        \centering
        \caption{Dataset for Open World Evaluation with Caching}
        \label{tab:dataowcache}
\begin{tabular}{lcccc}
        \toprule
        \tabhead{Type}    & \multicolumn{2}{c}{\tabhead{Training Data}}    &  \multicolumn{2}{c}{\tabhead{Test Data}}\\
        \cmidrule(l){2-3} \cmidrule(l){4-5}
        & \tabhead{Apps} & \tabhead{Traces p. App}    & \tabhead{Apps} & \tabhead{Traces p. App} \\
                \midrule
                Unknown                                &                -  & -         &        120 & 4                        \\
                Monitored                        &                20 & 8        &        20  & 2                        \\
                Unmonit.                        &                80 & 2         &        -   & -                        \\
                \bottomrule
\end{tabular}
\end{table}

\begin{figure}[tb]
	\centering
	\includegraphics[width=.95\linewidth]{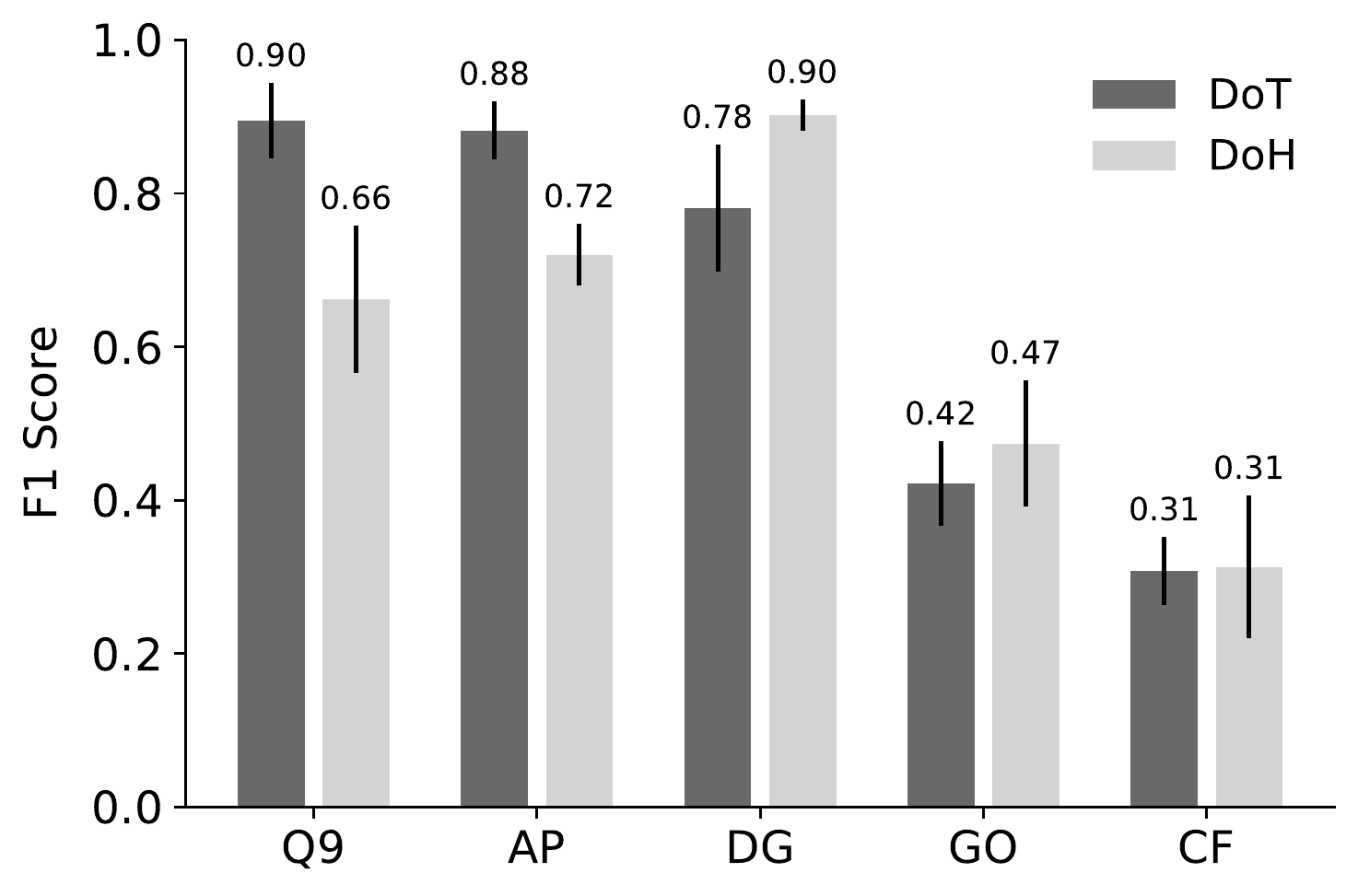}
	\caption{Macro-F1 Scores for Open World with Caching}
	\label{fig:owcache}
	\Description{Macro-F1 Scores for Open World with Caching for all selected resolver.}
\end{figure}

We report F1 scores for the multi-class case in Fig.~\ref{fig:owcache}. 
For DoT, \segram{} reaches F1 scores between 0.78 and 0.90 for all resolvers without padding.
If DoT resolvers apply padding, F1 scores drop to 0.42 for Google or 0.31 for Cloudflare.
For unpadded DoH traffic, \segram{} achieves F1 scores between 0.66 and 0.90.
If resolvers apply padding, \segram{} still reaches F1 scores of 0.47 for Google and 0.31 for Cloudflare.

These results are in line with expectations: Caching reduces the F1 scores by 0.05 to 0.23 in the open world setting. The results, nevertheless, indicate that \segram{} can still identify apps in practical environments.

\section{Prevalence of Padding}\label{sec:measurements}
As padding reduces the accuracy of traffic analysis attacks and is a recommended countermeasure, we evaluate how many recursive resolvers actually support padding.
According to RFC 7830, DNS responses must be padded if recursive resolvers receive a DNS query containing the padding option~\cite{mayrhofer2016ednspadding}.
As specified by RFC 8932, there are also specific recommendations for DoT and DoH resolvers to enable padding as protection mechanism against traffic analysis~\cite{dickinson2020recommendations}.
For our measurements, we selected multiple recursive resolvers from different sources---all of them with a focus on privacy.\footnote{\url{https://dnsprivacy.org/wiki/}, \url{https://www.privacytools.io/providers/dns/}, \url{https://github.com/curl/curl/wiki/DNS-over-HTTPS}}

For DoT, we extracted 56 resolvers from the aforementioned sources. 
To evaluate the padding behavior of the selected resolvers, we constructed padded DNS queries using the \textit{kdig} tool.
Table~\ref{tab:pad_summary} shows the results for the selected DoT resolvers. 
Only 64\,\% of the resolvers sent a valid response for our DNS query; for others, we received error messages or no response at all.
Out of these valid DoT responses, only 33\,\% were padded as recommended by RFC 8467~\cite{mayrhofer2018padding}.
Although DNS privacy service operators should consider padding~\cite{dickinson2020recommendations}, we did not receive any padding from 64\,\% of the resolvers.
Also, 3\,\% applied custom padding different from the recommendation.
Our measurements indicate that 64\,\% of the DoT resolvers in our sample ignore current recommendations and thus fail to support the EDNS0 padding option.

For DoH, we found 123 resolvers based on our selection.
To evaluate their padding strategies, we query the DoH resolvers with manually constructed DNS queries.
We check if the padding option is present in the response and determine which kind of padding is implemented. 
We received valid DNS responses from 77\,\%  of DoH resolvers. 
As shown in Table~\ref{tab:pad_summary}, 81\,\% of these DoH resolvers answered without any padding. 
Only 12\,\% of the DoH resolvers are padding DNS responses to multiples of 468 bytes. 
A minority of 7\,\% applies custom padding strategies. 
Overall, our measurements show that 81\,\% of the selected DoH resolvers do not follow the recommendations and respond without padding~\cite{dickinson2020recommendations}.
We have notified the affected operators before the publication of this paper to raise awareness about the padding option.
\begin{table}[tb]
	\centering
	\caption{Padding Strategies of Recursive Resolvers}
	\label{tab:pad_summary}
	\begin{tabular}{lcc}
		\toprule
		\tabhead{Answer}    &  \tabhead{DoT Resolver} & \tabhead{DoH Resolver}\\
		\midrule
		No Padding          &    0.64    &    0.81       \\    
		Custom Padding      &    0.03    &    0.07       \\
		EDNS 468 Padding    &    0.33    &    0.12       \\
		\bottomrule
	\end{tabular}
\end{table}

\section{Discussion}\label{sec:discussion}
Encrypting DNS messages is not enough to protect Android users' privacy against traffic analysis attacks.
Suggestions to apply padding to DNS queries and responses complicate those attacks but do not prevent them entirely.
If DNS responses are not padded, \segram{} identifies Android apps with accuracies between 94\,\% and 99\,\%, depending on the resolver.
With padding, \segram{} is able to correctly identify 66–72\,\% of the apps for DoT and 64–72\,\% for DoH.

The related work~\cite{houser2019investigation,siby2020encrypted,bushart2020padding} neglected mobile clients and considered only website fingerprinting based on encrypted DNS traffic.
For a valid comparison with \segram{}'s accuracy, we reimplemented several approaches from website fingerprinting to assess their effectiveness for app identification.
In comparison, \segram{} reaches similar accuracy values as other traffic analysis attacks without padding.
For padded DoT traffic, the accuracy of \segram{} is on the same level as state of the art traffic analysis attacks on encrypted and padded DNS traffic~\cite{bushart2020padding}.
On Cloudflare's DoT resolver, \segram{} reaches the same accuracy as the B\&R attack (0.72).
For Google's DoT resolver, \segram{}'s accuracy is lower with 0.67 (--0.11).
However, in comparison to the B\&R attack, \segram{} is improving the runtime for the classification substantially.
For Google's DoT resolver, \segram{} is able to classify 100 traffic traces in 0.04 seconds on average while the B\&R attack needs 46 seconds on average.

For padded DoH traffic, \segram{} outperforms the B\&R attack by 5–8 percentage points.
Additionally, \segram{} is also improving the runtime on DoH traffic, i.\,e., the runtime is similar on DoT and DoH traffic.
For instance, our measurements have shown that the B\&R attack needs 212 seconds to classify 100 traffic traces for Google's DoH resolver.
In contrast, \segram{} is able to classify these 100 traces in 0.04 seconds.
Given these findings, we believe that \segram{} is more suitable for practical scenarios because calculations for the kNN classifier with the Damerau-Levenshtein distance are computationally expensive.

Furthermore, we considered several influencing factors such as the sample size, the influence of updates, and the choice of the recursive resolver, which might lower \segram{}'s accuracy.
Traffic analysis attacks will be easier to carry out if an adversary (1) needs only a few traffic traces, (2) does not need to retrain the model often, and (3) is able to use existing data from one resolver to identify apps from other resolvers.
We have shown that a relatively small number of traces is sufficient to identify most of the apps with high accuracy.
Besides, we have shown that the influence of updates depends strongly on whether the resolver supports padding.
With padding, the accuracy decreased substantially after three weeks, while without padding, the accuracy remains nearly constant despite app updates.

Training on one resolver and predicting on another resolver seems challenging.
However, with Google's and Cloudflare's DoT resolver, \segram{} still achieves accuracies of up to 41\,\% whereas a random classifier would reach an accuracy of 0.85\,\% on 118 apps.
This result  suggests that using the recommended padding strategy might lead to a counterintuitive situation: traces obtained from different resolvers become more similar.
Consequently, the adversary's effort for data collection decreases.

Fingerprinting attacks are often criticized for not being applicable to the real world.
Therefore, we evaluated \segram{} under more realistic conditions with enabled caching and in the open world.
Our results indicate that caching effects are relatively small for most resolvers.
Although this is in line with previous results from website fingerprinting, it might be that our experimental setup has influenced the results.
While we observe on average a fewer number of DoT packets with caching (27 vs. 32), we did not observe the same effect for DoH.
Moreover, if we would have included traces of subsequent app starts of the same app, accuracy values might decrease stronger, i.\,e., we would not be able to detect all app starts.
However, we should still be able to detect the first one.

Our open world evaluation shows that \segram{} is able to differentiate between a small set of known apps and a larger set of unknown apps.
If there is no padding, we reach F1 scores of 0.92 for DoT and 0.85 for DoH, respectively.

While padding might be somewhat effective against \segram{}, future attacks might improve accuracy values even further.
However, we have also shown that many resolvers do follow the recommendations to support the EDNS0 padding option, which indicates missing awareness regarding traffic analysis attacks.
Given the resolvers in our experiment, privacy-aware users have the difficult choice between resolvers with padding operated by Google and Cloudflare or “privacy-aware” resolvers operated by non-profit organizations that \emph{do not support} padding, i.\,e., are more susceptible to traffic analysis attacks.

\paragraph{Limitations}
Our experiment has the following limitations:
Firstly, network traffic was routed through a VPN, which might have affected our results.
However, by logarithmizing the inter-arrival times between the DoH/DoT packets, n-grams of DNS sequences might not be that susceptible to changing network conditions.
Future research might evaluate \segram{} in different networks.
Secondly, the network itself was limited to IPv4 traffic.
With a dual stack, i.\,e., with IPv4 and IPv6, the number of exchanged DNS packets increases and would provide the classifier with more information, possibly leading to better classification results.
Thirdly, caching is hard to evaluate reproducibly and realistically at the same time.
Even in the warm cache scenario, our data collection did not consider different possible states of the cache resulting from different orders of app starts or usage.
Specifically, we did not consider subsequent app starts.

\section{Conclusion}\label{sec:conclusion}
We propose \segram{}, a novel traffic analysis attack to identify Android apps using solely DoT or DoH traffic.
We compare its effectiveness with other state-of-the-art traffic analysis attacks.
For resolvers without padding, \segram{} and previous attacks reach accuracy values as high as 99\,\%.
For resolvers with padding, \segram{} reaches an accuracy of up to 72\,\% for DoT; for DoH, our attack has an accuracy of 72\,\%, outperforming the other attacks by up to 8 percentage points.
Our runtime benchmarks indicate that the computational effort of \segram{} is low enough to fingerprint a large number of devices in practice.

We also show that \segram{} performs reasonably well in more realistic conditions:
Firstly, DNS caching has only a negligible influence on accuracy in the closed world setting.
Secondly, while app updates can decrease accuracy substantially, adversaries can re-train their classifier with small sample sizes.
Finally, in the open world setting, when adversaries have to deal with unknown apps, \segram{} still achieves a recall of 0.92 at a precision of 0.93 without padding.
When messages are padded, the recall drops to 0.56 at a precision of 0.43 (given values apply to DoT).
According to these findings, app fingerprinting is still possible and DoT and DoH provide less privacy than expected.

Moreover, we analyze the prevalence of padding in the wild and find that the majority of the tested resolvers does not implement padding (DoT: 64\,\%, DoH: 81\,\%).
Quite worryingly, many privacy-focused resolvers operated by non-profit organizations do not support padding, while large companies such as Google and Cloudflare do support padding.
Besides standardizing more effective padding approaches, the community should look into increasing awareness about this privacy feature among resolver operators.

\section*{Acknowledgements}
This work has been funded by the German Federal Ministry of Education and Research (BMBF) as part of the BMBF project WINTERMUTE (16KIS1128).

\bibliographystyle{ACM-Reference-Format}
\bibliography{bibliography}

\end{document}